\documentclass[aps,prb,twocolumn,groupedaddress,showpacs]{revtex4}
\usepackage{graphicx}
\usepackage{isolatin1}
\newcommand{\myscalebox}[1]{\scalebox{0.42}[0.42]{#1}}
\begin{document}
\title{No spin-glass transition in the   ``mobile-bond'' model }
\author{A. K. Hartmann}
\affiliation{Institut f\"ur Theoretische Physik, Universit\"at G\"ottingen,
Bunsenstra\ss{}e 9, 37037 G\"{o}ttingen, Germany}
\date{\today}
\begin{abstract}
The recently
introduced ``mobile-bond'' model for two-dimensional
spin glasses is studied. The model is characterized by an annealing
temperature $T_{\rm q}$. On the basis of Monte Carlo simulations of
small systems it has been claimed that this model exhibits a
non-trivial spin-glass transition at finite temperature for small
values of $T_{\rm q}$.

Here the model is studied by means of exact ground-state calculations
of large systems up to $N=256^2$. The scaling of domain-wall energies is
investigated as a function of the system size. For small values $T_{\rm q}<
0.95$ the system behaves like a (gauge-transformed) ferromagnet having a
small fraction of frustrated plaquettes. For $T_{\rm q}\ge 0.95$ the
system behaves like the standard two-dimensional 
$\pm J$ spin-glass, i.e.\ it does {\em not}
exhibit a phase transition at $T>0$.
\end{abstract}
\maketitle

Spin glasses \cite{reviewSG} are the prototype model for disordered systems 
investigated extensively  during the last three decades in
statistical physics. 
These systems exhibit complex energy landscapes
resulting in many interesting phenomena like
glassy behavior  and aging. Despite much effort, still many open
questions exists.
The question about the lower critical dimension of Ising spin glasses
had been discussed for quite a while
\cite{mcmillan1984,bray1984,shirakura1997,matsubara1998}. 
Now it is clear that in two
dimensions non stable spin-glass phase at finite temperature exists
\cite{rieger1996,kawashima1997,stiff2d,houdayer2001,carter2002}.
This has motivated the search for other two-dimensional
spin-glass-like systems exhibiting a $T_{\rm c}>0$ \cite{alex-rcf2001}. Recently 
D.K. Sunko has proposed \cite{sunko2002} a  ``mobile-bond''
model where quenched-disorder realizations
of $\pm J$ spin glasses are created by an annealed simulation,
allowing the bonds to move. The system is equilibrated at high
temperature,  followed by a quench to a
temperature $T_{\rm q}$. Sunko has performed Monte-Carlo simulations of
systems up to size $L=16$ and claimed that for low quenching temperatures
$T_{\rm q}$ the model exhibits a spin-glass transition at finite
temperature $T_{\rm c}>0$.

In this rapid communication,
the model is studied by means of exact ground-state calculations
of large systems up to $L=256$. The scaling of domain-wall energies
\cite{banavar1982,bray1984,mcmillan1984} is
studied as a function of the system size. It is shown here that the model
exhibits no spin-glass transition at finite temperature. For small
values of $T_{\rm q}<0.95$ the system exhibits ferromagnetic order,
while at $T_{\rm q}$ a transition to the normal two-dimensional
spin-glass behavior is found, i.e. $T_{\rm c}=0$.

The model consists  of $N=L^2$ Ising spins $S_i=\pm 1$ on a square
lattice with the Hamiltonian
\begin{equation}
{\cal H} = -\sum_{\langle i,j \rangle} J_{ij} S_iS_j\,,\label{eq:hamil}
\end{equation}
where the sum runs over all pairs of nearest neighbors $\langle i,j \rangle$
and the $J_{ij}=\pm J$ are quenched random variables.

The realizations are prepared exactly in the same way as in
Ref. \onlinecite{sunko2002}. For each realization,
first $N$ bonds with strength $+J$ and  $N$ bonds with strength $-J$ 
are distributed randomly among all  $2N$
bonds. Then the values of all spins are
set randomly to orientations $S_i=\pm 1$. Next, an annealed Monte-Carlo (MC)
simulation \cite{STAT-mc2} 
is performed. This means, at each step either a spin is allowed to flip
or two bonds incident to the same site are allowed to exchange their
positions. Each choice occurs with probability 0.5. Each step is
accepted with the usual Metropolis probability depending on the energy change 
according to the Hamiltonian (\ref{eq:hamil}). First, the system is
equilibrated at high temperature $T=5$ for 1000 MC steps per 
spin  (MCS) \cite{footnote1}. Finally the system is quenched to $T=T_{\rm q}$
and simulated for further 1000 MCS. The result is a realization of the
disorder which can used for further treatment, here ground-state calculations
are applied.

In greater than two dimensions, or in the presence of a magnetic field,
the exact calculation of spin-glass ground states belongs
to the class of NP-hard problems \cite{barahona1982,opt-phys2001}. 
This means that
only algorithms with exponentially increasing running time are known.
However, for the special case of a planar system without magnetic field, e.g.
a square lattice with periodic boundary conditions in at most one direction,
there are
efficient polynomial-time ``matching'' algorithms \cite{bieche1980}. The
basic idea is to represent each realization of the disorder by its frustrated
plaquettes \cite{toulouse1977}. Pairs of
frustrated plaquettes are connected by paths
in the lattice and the weight of a path is defined by the sum of the absolute
values of the coupling constants which are crossed by the path.
A ground state corresponds the set of paths with minimum total
weight, such that each frustrated plaquette is connected to exactly one
other frustrated plaquette. This is called a minimum-weight perfect
matching.
The bonds which are crossed by paths
connecting the frustrated plaquettes are unsatisfied in 
the ground state, and all other bonds are satisfied.

For the calculation of the minimum-weight perfect matching, efficient
polynomial-time algorithms are available \cite{barahona1982b,derigs1991}.
Recently, an implementation has been presented\cite{palmer1999}, where 
ground-state energies of large systems of size $N\le 1800^2$ were calculated. 
Here, an algorithm from the LEDA library \cite{leda1999} has been
applied, which allows a quick implementation. It was not necessary to
go beyond $N=256^2$ (with is much larger than $N=16^2$ in the
original work \cite{sunko2002}) to obtain reliable results.

To study whether an ordered phase is stable at finite temperatures,
the following procedure is usually applied 
\cite{mcmillan1984,bray1984,rieger1996,kawashima1997,%
alex-stiff,alex-4d,alex-rcf2001}.
First a ground state 
of the system is calculated.
Then the system is perturbed to introduce a domain wall and the
new ground-state energy is evaluated. Typically, the
system initially has periodic boundary conditions in both directions, and the
perturbation involves replacing periodic by antiperiodic boundary
conditions in one direction. The domain-wall energy $\Delta E$ 
is given by the difference
of the two ground-state energies. In case the model exhibits long-range
ferromagnetic order at non-zero temperatures, 
the domain-wall energy, averaged over many independent
samples,  has to increase with system size.
E.g. for a pure 2d ferromagnet, the domain wall consists of a straight line,
resulting in $\Delta E \sim L$. For a spin-glass, none of the ground states
with periodic and antiperiodic boundary conditions has a priori a lower
energy. Hence one studies the absolute value of the domain-wall energy
 to detect whether the systems exhibits 
spin-glass ordering at finite temperatures. 

However we cannot apply the matching algorithm
for boundary conditions which ``wrap around'' in both directions. For this
reason, here the periodic boundary conditions in the
y-direction are broken for each realization (by setting the bonds
connecting the first and the last row to zero). 
This has no influence on the fact of whether the systems orders or not
 because the change of the boundary conditions 
to create the domain walls occurs
in the x-direction perpendicular to the open boundaries.

Here system sizes $L=4,6,8,\ldots
192, 256$ are considered. For each size, 1000 independent realizations of the 
disorder were generated for quenching temperatures
$T_{\rm q}=0.1,0.9,0.95,1.0$ and $1.5$. Then
ground states with periodic (P) and antiperiodic (AP)
boundary conditions in $x$-direction were calculated using the exact
matching algorithm, resulting in ground-state energies $E^0_{\rm P}$
resp.\ $E^0_{\rm AP}$. The change in the boundary conditions introduces a
domain wall in each realization 
with energy $\Delta E=E^0_{\rm AP}-E^0_{\rm P}$.

\begin{figure}[htb]
\begin{center}
\myscalebox{\includegraphics{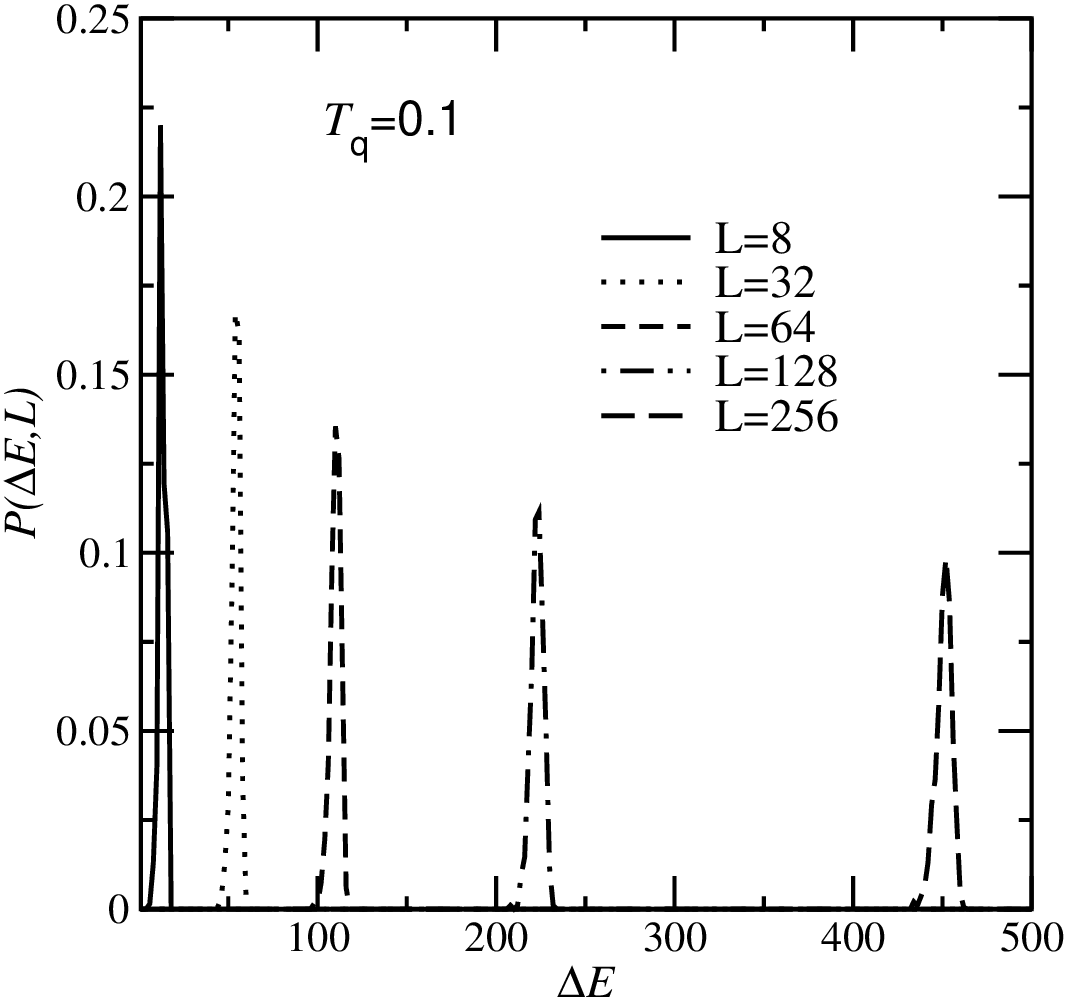}}
\end{center}
\caption{Distribution $P(\Delta E,L)$ 
of domain-wall energies for $T_{\rm q}=0.1$ and
  system sizes $L=8,32,64,128,256$.}
\label{figAnnealed01stat}
\end{figure}

\begin{figure}[htb]
\begin{center}
\myscalebox{\includegraphics{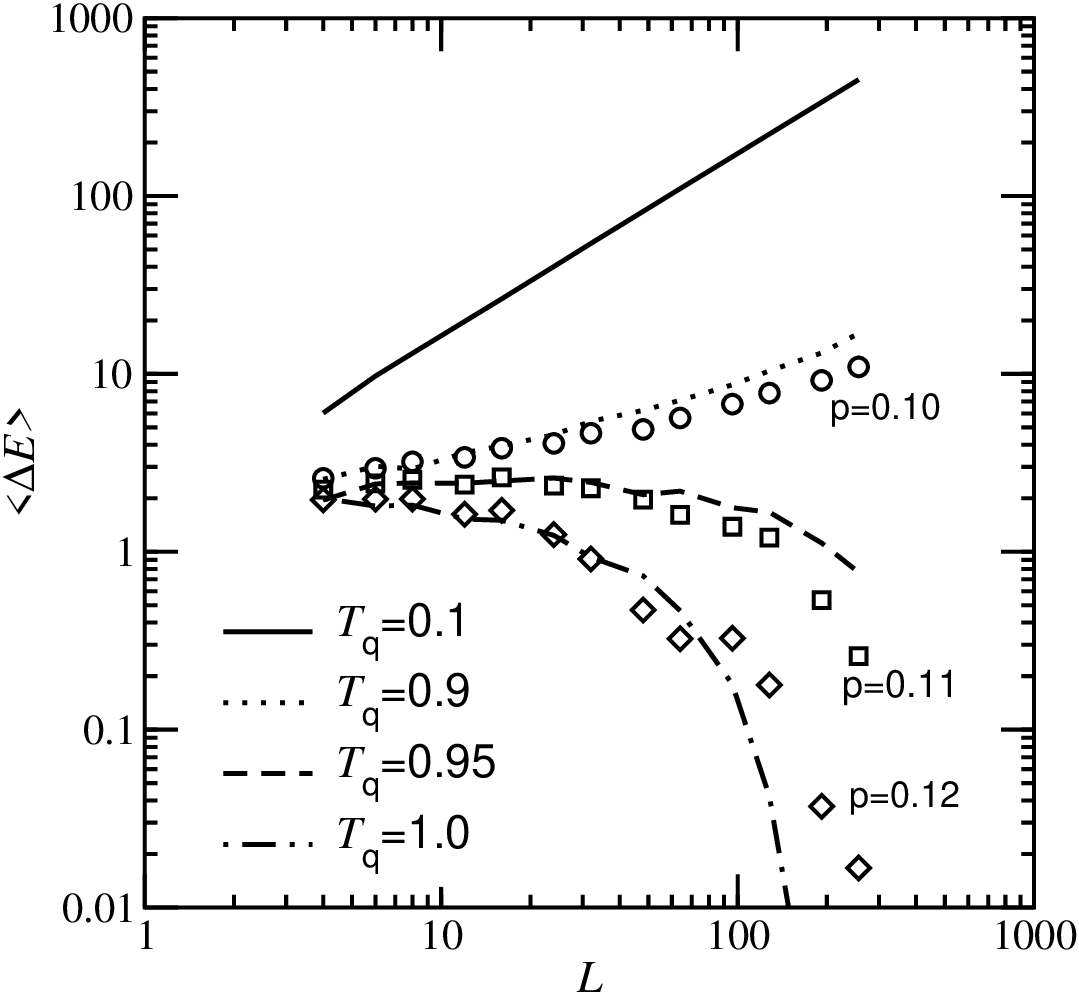}}
\end{center}
\caption{Mean value of the domain-wall energy $\langle \Delta E
  \rangle$ as a function of system
  size for $T_{\rm q}=0.1, 0.9, 0.95, 1.0$ (displayed by lines). 
Also the domain-wall
  energy for the $\pm J$ random-bond model with concentrations
  $p=0.1,0.11,0.12$ of the antiferromagnetic bonds is shown (symbols).}
\label{figAnnStiff0}
\end{figure}

First, we consider a very low quenching temperature $T_{\rm q}=0.1$, which
was claimed in Ref. \onlinecite{sunko2002} to exhibit a spin-glass 
transition at $T_{\rm c}/J=2.22(1)$ and no ferromagnetic order. In Fig.\
\ref{figAnnealed01stat} the distribition $P(\Delta E, L)$
over the disorder of the domain-wall energies is displayed for
different system sizes $L$. Clearly, the domain-wall energies grow
strongly with system size, which is an indicator for ferromagnetic
order. Below (c.f. Fig. \ref{figAnnStiff0}) it is shown that indeed
the disorder average $\langle \Delta E \rangle$ grows linearly with
$L$, as in the normal ferromagnet. But the model exhibits
no global magnetic moment, as found already in Ref. \onlinecite{sunko2002}.
 This is due to fact
that 50\% of all bonds are antiferromagnetic. 
Nevertheless, the model behaves like a ferromagnet. 
The reason is that the bonds are distributed in
the system such that only few frustrated plaquettes are present.
Hence, each realization can be mapped via a local gauge transformation
on a ferromagnet with a small number of antiferromagnetic bonds. This
explains the fact that in Ref. \onlinecite{sunko2002} the critical
exponent of the correlation length found at $T_{\rm c}/J=2.22$ was 
indeed that of the pure ferromagnet.

\begin{figure}[htb]
\begin{center}
\myscalebox{\includegraphics{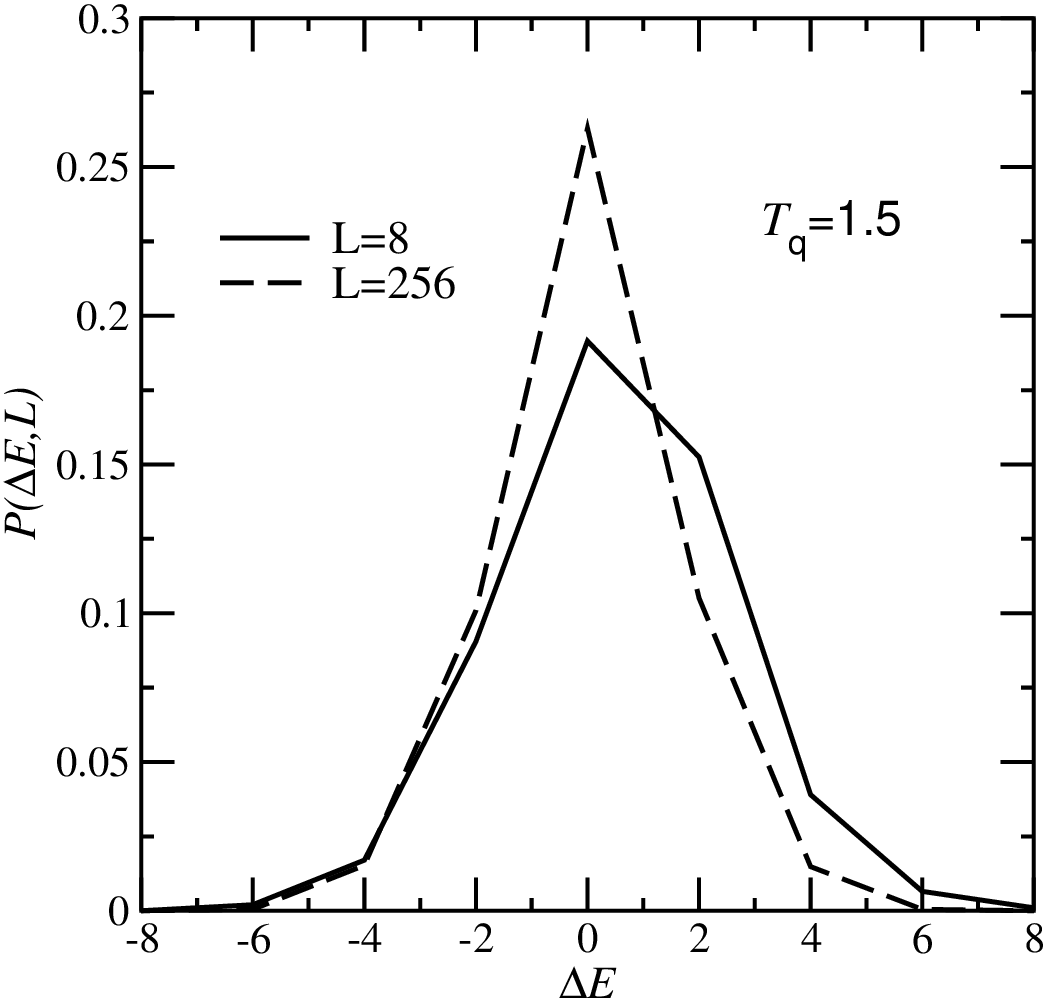}}
\end{center}
\caption{Distribution of domain-wall energies for $T_{\rm q}=1.5$ and
  system sizes $L=8,256$.}
\label{figAnnealed15stat}
\end{figure}

Next, a large quenching temperature 
$T_{\rm q}=1.5$ is considered. In Fig. \ref{figAnnealed15stat}
again the distribution of domain-wall energies for different sizes are
show. For large sizes, 
the distrubtions are centered around $\Delta E=0$ indicating the
absence of ferromagnetic order. Furthermore, the width of the
distributions decreases slightly with increasing system size, which
shows that spin-glass order is not stable against thermal fluctuations. 
This is the usual situation found for the 
two-dimensional $\pm J$ spin glass \cite{kawashima1997,stiff2d} (having
$T_{\rm c}/J=0$).
 
\begin{figure}[htb]
\begin{center}
\myscalebox{\includegraphics{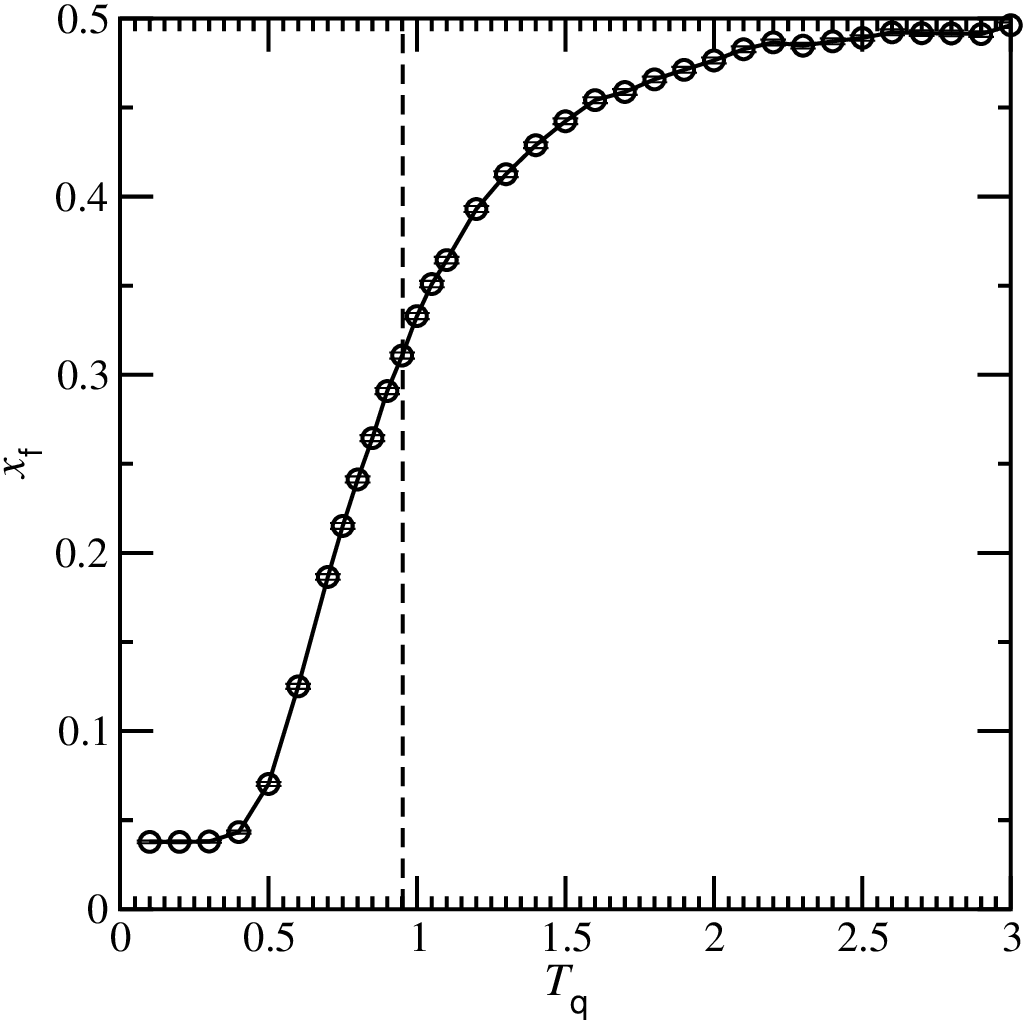}}
\end{center}
\caption{Fraction of frustrated plaquettes present after the quench to
  temperature $T_{\rm q}$ as a function of $T_{\rm q}$. The line is a guide to
  the eyes only.}
\label{figFracFrustAnnealed}
\end{figure}

To understand the behavior of the system better, next it is analysed
as a function of $T_{\rm q}$. The behavior is probably  mainly
determined by the fraction of frustrated plauqettes.  
In Fig.\ \ref{figFracFrustAnnealed} the average fraction of frustrated
plaquettes of the quenched realizations is shown as a function of
$T_{\rm q}$. This can be compared with the standard $\pm J$ random
bond model (with parameter $p\in [0,1]$),
which has on average $2Np$ antiferromagnetic and $2N(1-p)$
ferromagnetic bonds. This results in a average fraction 

\begin{equation}
x_{\rm f}=4x(1-x)[x^2+(1-x)^2]
\end{equation}
of frustrated plaquettes \cite{kirkpatrick1977}. For the $\pm J$ model,
a ferromagnet spin-glass ($T_{\rm c}=0$) transition
occurs \cite{grinstein1979,freund1989,bendisch1992,kawashima1997,merz2002} 
near $p=0.11$ were $x_f(0.11)\approx 0.31$. This corresponds to a
quenching  temperature $T_{\rm q}\approx0.95$, see Fig.\
\ref{figFracFrustAnnealed}. Hence, for a comparison, simulations near
$p=0.11$ for the $\pm J$ model and near $T_{\rm q}\approx 0.95$ for
the ``mobile-bond'' model have been performed. Furthermore both models were
investigated for two other pairs of parameters exhibiting similar
concentrations of frustrated plaquettes: $T_{\rm q}=0.9;\,p=0.1$
and $T_{\rm q}=1.0;\,p=0.12$.

In Fig.\ \ref{figAnnStiff0} the mean value  $\langle \Delta E
  \rangle$ of the domain-wall energy is shown as a function of the system
  size for $T_{\rm q}=0.1, 0.9, 0.95, 1.0$. 
For $T_{\rm q}=0.1$ a
  clear linear increase occurs, corresponding to a normal
  ferromagnet. For $T_{\rm q}=0.9$ the domain-wall energy still
  increases with system size. The resulting values are very similar to
  the domain-wall energies found at $p=0.1$ for the  $\pm J$ model.
  For $T_{\rm q}=0.95$ and $T_{\rm q}=1$ the mean domain-wall energies
  decrease as a function of the system size, hence no ferromagnetic
  order persists. In these cases the data is almost equal to the
  results for $p=0.11$ resp. $p=0.12$ of the $\pm J$ model. 
  Please note that it is not claimed here that e.g. $T_{\rm q}=0.9$
  corresponds exactly to $p=0.1$. But it seems certainly possible
  to chose $p$ such that the results for both models agree exactly.

\begin{figure}[htb]
\begin{center}
\myscalebox{\includegraphics{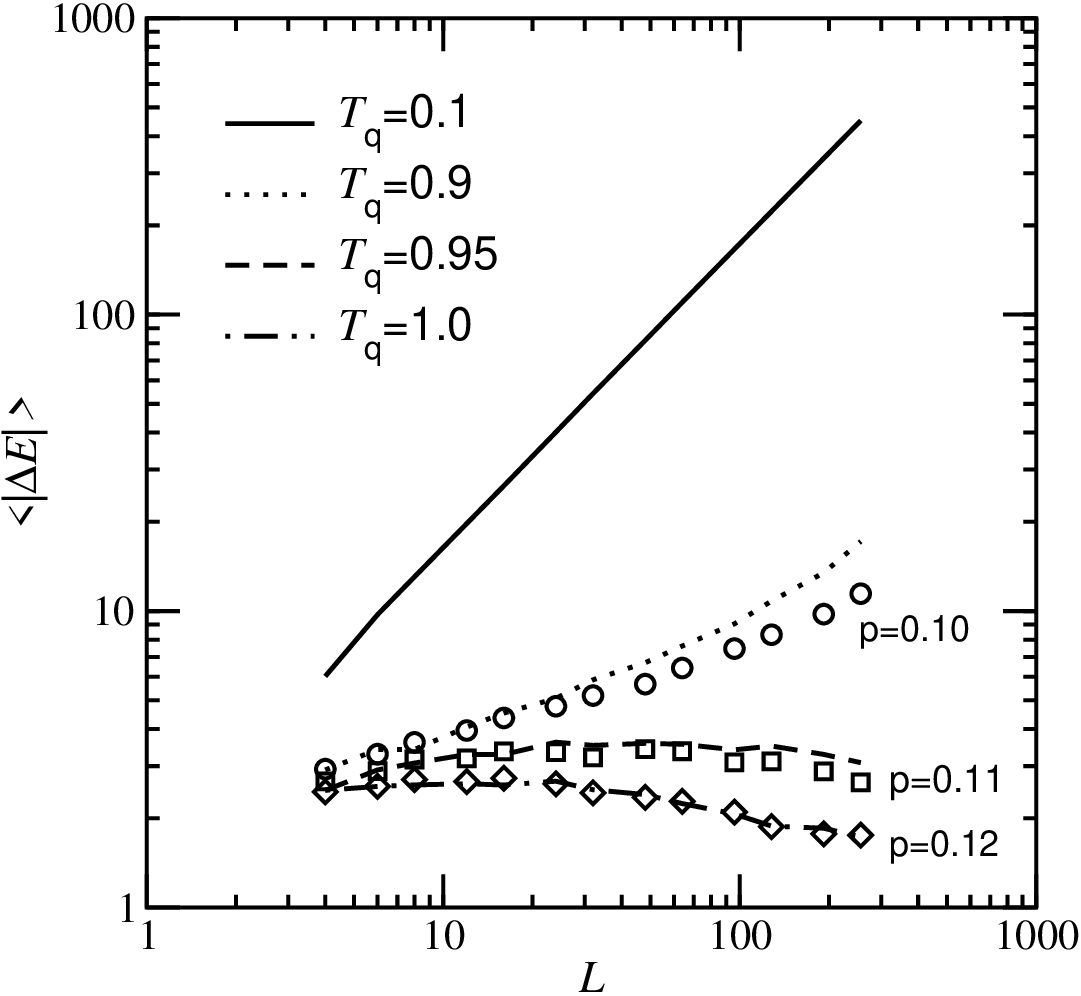}}
\end{center}
\caption{Mean absolute value of the domain-wall energy $\langle |\Delta E|
  \rangle$ as a function of system
  size for $T_{\rm q}=0.1, 0.9, 0.95, 1.0$ (displayed by lines). 
Also the absolute value of  the domain-wall
  energy for the $\pm J$ random-bond model with concentrations
  $p=0.1,0.11,0.12$ of the antiferromagnetic bonds is shown (symbols).}
\label{figAnnStiff1}
\end{figure}

In Fig.\ \ref{figAnnStiff1} the corresponding results for 
the mean $\langle |\Delta E|\rangle$ of
the absolute value of the domain-wall energy is shown. For values $T_{\rm
  q}\le 0.9$ again an increase is observed, due to the increase of the
mean (non absolute) $\langle \Delta E \rangle$. If spin-glass ordering
existed in a system, then  $\langle |\Delta E|\rangle$ would {\em increase}
with growing $L$, while $\langle \Delta E\rangle$ has to decrease. 
For $T_{\rm q}=0.95,1.0$ $\langle |\Delta E|\rangle$ increases only for small
system sizes (which may cause signs of a stable spin-glass phase
when simulating only small systems), while it starts to {\em decrease}
with $L$ for larger values of $L$. 
 Hence spin-glass order is destroyed for any finite temperature $T>0$.
Please note that again the results at $T_{\rm q}=0.9,0.95,1.0$ agree
well with the results at $p=0.10,0.11,0.12$ for the $\pm J$ model.

To conclude, in this work the recently proposed
``mobile-bond'' spin-glass model has been investigated. An exact
ground-state matching algorithm has been applied, allowing to study
large system sizes like $N=256^2$. The model turns out to be mainly
equivalent to the $\pm J$ random-bond model, which has been studied
extensively in the past. 
Hence, for low values of the annealing
temperature $T_{\rm q}$, the model (corresponding to small
concentrations $p$ of the antiferromagnetic bonds in the $\pm J$ model)
exhibits ferromagnetic order. The only difference is that the Sunko
model exhibits no magnetic moment, since by construction the number of
ferromagnetic bonds equals the number of antiferromagnetic bonds. Both
models can be mapped onto each other by local gauge-transformations,
the characteristic parameter is the fraction of frustrated plaquettes.

For larger values of $T_{\rm q}\ge 0.95$ (corresponding to $p\ge 0.11$)
the model displays the standard behavior of a 
two-dimensional $\pm J$ spin-glass, hence no order for
$T>0$ exists. To summarize, the ``mobile-bond'' model does {\em not}
exhibit a finite-temperature spin-glass transition at any value of $T_{\rm
  q}$,  opposed to the claims made in Ref. \cite{sunko2002}.

{\em Acknowledgements:} The author obtained financial support from the
{\em VolkswagenStiftung} (Germany) within the program
``Nachwuchsgruppen an Universitäten''. He thanks A.P. Young for
critically reading the manuscript.
The simulations were performed on a Linux
cluster at the {\em Gesellschaft für wissenschaftliche
  Datenverarbeitung Göttingen} (GWDG) (Germany).


\begin{thebibliography}{99}
\bibitem{footnote1} The time to equilibrate the system is only few MC
  steps per spin. This has been checked by starting with fully
  frustrated and fully unfrustrated systems and measuring the level of
  frustration and the energy as a function of the number of MC steps.
\end{thebibliography}
\end{document}